\documentclass[conference]{IEEEtran}
\IEEEoverridecommandlockouts

\usepackage{cite}
\usepackage{amsmath,amssymb,amsfonts}
\usepackage{graphicx}
\usepackage{textcomp}
\usepackage{xcolor}
\usepackage{booktabs}
\usepackage{array}
\usepackage{algorithm}
\usepackage{algpseudocode}
\usepackage{multirow}
\usepackage[normalem]{ulem}
\usepackage{url}

\def\BibTeX{{\rm B\kern-.05em{\sc i\kern-.025em b}\kern-.08em
    T\kern-.1667em\lower.7ex\hbox{E}\kern-.125emX}}

\begin{document}

\title{Hybrid spatial–temporal graph neural network Powered NDTs: Towards Next-Gen Smart Infrastructure Twins}

\author{\IEEEauthorblockN{John Sengendo}
\IEEEauthorblockA{C.N.I.T and \\
University of Trento, ITALY.\\
Email: john.sengendo@unitn.it}
\and
\IEEEauthorblockN{Fabrizio Granelli}
\IEEEauthorblockA{C.N.I.T and \\
University of Trento, ITALY.\\
Email: fabrizio.granelli@unitn.it}

}

\maketitle

\begin{abstract}
Network Digital Twins (NDTs) enable proactive network management and optimization by predicting system behavior before control actions are applied to live infrastructures, supporting critical operations in Internet Service Provider (ISP) networks and wide-area networks (WANs). However, to anchor the superior performance NDTs promise to provide, key enabler techniques are required. Given that mobile networks are modeled as graphs, graph-based architectures such as graph neural networks (GNNs) have shown promising performance in modeling network behavior. This work proposes a novel Hybrid Spatial-Temporal Graph Neural Network (HSTGNN) architecture. Unlike single-branch GNN approaches, we propose a multi-scale design that combines three complementary message-passing paradigms: local neighborhood aggregation, spectral filtering, and learnable attention-based weighting. When benchmarked against other approaches, the proposed HSTGNN achieved superior performance delivering a coefficient of determination score of approximately 0.8816, 17.5\% better than the best baseline ChebNet. Furthermore, HSTGNN achieved the lowest Mean Absolute Error (MAE) of 0.0300, and Root Mean Squared Error (RMSE) of 0.0458, significantly outperforming baseline frameworks and certifying the proposed framework's capability in enabling NDTs.

\end{abstract}


\section{Introduction}
Future mobile networks require predictive and higher automation mechanisms \cite{vaishnavi}, that can anticipate network behavior before changes are deployed in live operations. This need is becoming even more pressing as modern networks grow in scale, heterogeneity, and service sensitivity. Operators are no longer managing only static connectivity, they are managing latency-sensitive applications, dynamic traffic demands, geographically distributed datacenters, and increasingly complex control policies. As mentioned my authors in \cite{Rustamov}, traditional network management approaches are not well-suited for these tasks. Moreover, in such heterogenuous settings, an error in routing, power blackout, capacity planning, or congestion mitigation can quickly propagate into degraded quality of service and significant economic loss \cite{Huang2008ManagingCN}.

Network Digital Twins (NDTs) are seen as promising enablers to solve most of these challenges and enabling 6G promises underscored in \cite{fitzek2026promise}, through their capability of providing a virtual representation of the network and enabling scenario analysis, validation, and performance forecasting before actions are applied to the physical infrastructure, thus mitigating negative impact on network operations \cite{Poorzare}. An efficient NDT should not only mirror the topology, it should also support reasoning on the impact of the topology, connectivity, and state variation on operational metrics such as Round-trip time (RTT) and packet loss, as service level agreements (SLAs) depend on the capability to measure and monitor those metrics, as further noted in the protocol mechanism in \cite{Frost2011PacketLA}. This idea is in line with the Graph Neural Networks (GNNs) literature, where frameworks such as RouteNet presented by authors in \cite{rusek} are designed to capture the complex interactions between network topology, routing, and traffic patterns in order to predict metrics such as delay and jitter \cite{rusek}.
Additionally, because the network itself is represented as a graph, GNNs are key enablers in modeling network operations. Nodes correspond to network entities or measurement points, while edges encode structural relationships and connectivity constraints. Compared with traditional machine learning models that assume independent and identically distributed (IID) samples and operate solely on feature vectors \cite{zhou2020graph},graph neural networks can exploit both node attributes and the relational behavior among nodes. This enables more effective modeling of interdependencies that could be ignored in feature-based approaches.

\begin{figure}[htbp]
\centerline{\includegraphics[width=\columnwidth, height=3.7cm]{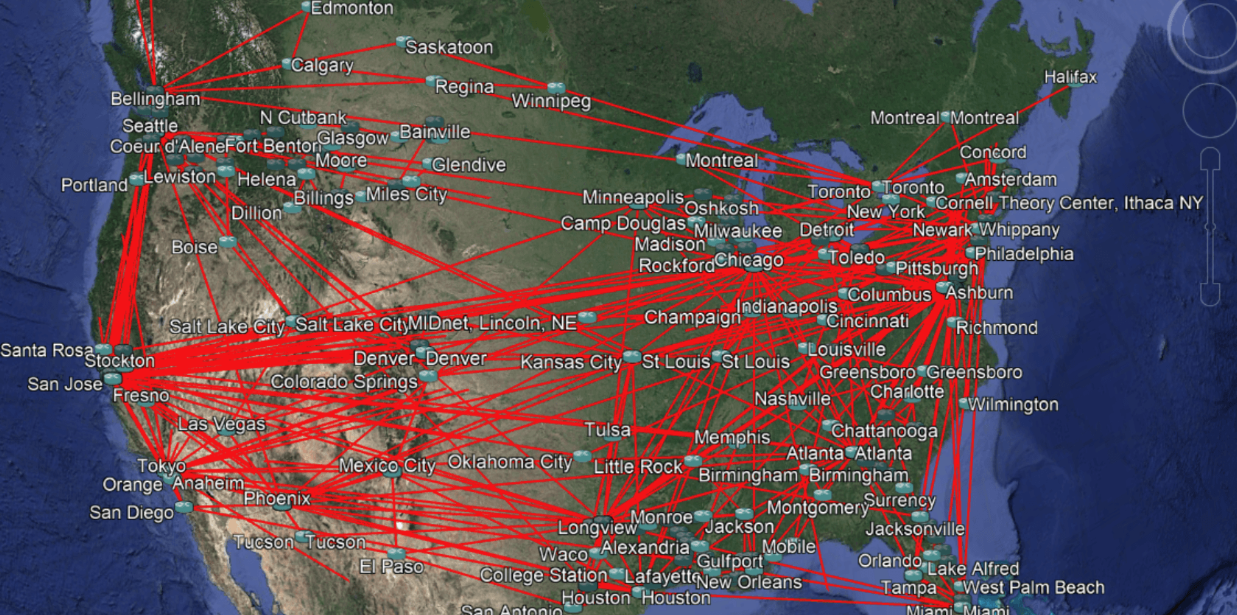}}
\caption{Visualization of connectivity on real-world network topologies \cite{afourmy3dzoo}.}
\label{fig:zoo}
\end{figure}

\begin{figure*}
\centerline{\includegraphics[width=\columnwidth,height=3.9cm,keepaspectratio]{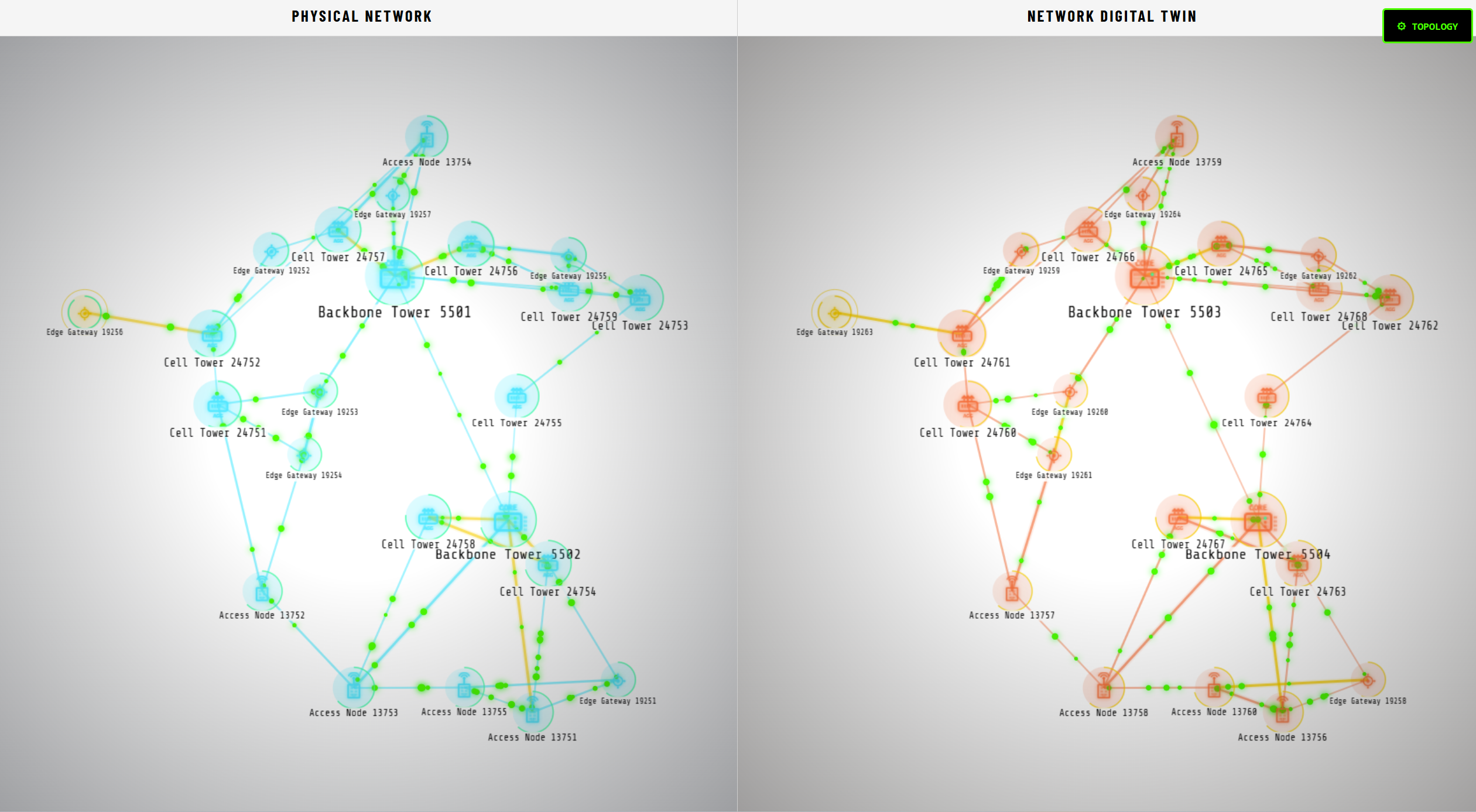}}
\caption{Graph-based representation of a physical network (left) and its digital replica (right) reflecting nodes and edges connectivity.}
\label{fig:topology}
\end{figure*}

A technical implementation presented by authors in \cite{Zacarias2025} evaluates four GNN architectures for the effectiveness of NDTs using RIPE Atlas data, and show that GraphTransformers are the strongest baseline in that setting. Their study is an important reference point because it demonstrates that graph learning is effective in developing  NDTs \cite{Zacarias2025}. However, most of the findings as demonstrated in the previous one mentioned mainly focus on comparing separate baseline architectures. In contrast, this paper proposes a hybrid model, HSTGNN, that integrates the strengths of multiple graph learning paradigms into a single architecture.
Additionally, to bring real-world realism into our framework, were we use a real world ISP dataset from Internet Topology Zoo \cite{afourmy3dzoo} to build a topology-aware NDT environment. A sample connectivity from the repository is shown in Figure~\ref{fig:zoo}, which represents connectivity in in North America. Additionally, for benchmarking purposes, our approach compares the proposed model framework against baselines GraphSAGE, ChebNet, ResGatedGCN, and GraphTransformer.

The rest of the paper is organized as follows. Section II reviews related works, Section III outlines the methodology, Section IV describes the HSTGNN architecture, Section V presents results and Section VI concludes with future directions.

\section{Related Work}
Graph Neural Networks (GNNs) have emerged as a powerful paradigm for modeling complex systems with relational structure, particularly in communication and networked environments. For example, a survey on these paradigms aiding intelligent modeling and orchestration is presented in \cite{tam2022gnn}, where authors survey and discuss their roles in networks. Moreover, in their discussion, the authors additionally highlight that GNNs effectively capture both local and global dependencies in graph-structured data, enabling improved prediction accuracy for network-level tasks such as routing, performance estimation, and resource optimization \cite{tam2022gnn}. Overlapping into the Network Digital Twins (NDTs) domain, recent practical work presented by authors in \cite{Zacarias2025} discusses different GNN architectures, evaluating their effectiveness for Digital Twins (DTs). Their study underpins how GNNs are key for the effectiveness of NDTs and benchmarks 4 GNN frameworks. Still intersecting with NDTs, studies show that GNNs and NDTs can be utilized to boost network performance, streamline routing, support network slicing, and management, as the works presented by the authors in \cite{9310275} suggests, where they exploit a novel graph neural network model on non-Euclidean graph structures.
Additionally, the fact that GNNs have been widely applied in learning and modeling graph-structured data as underscored by works in \cite{rusek}, this strength can be leveraged in networks by learning dependencies between nodes and links, outperforming traditional heuristics in optimizing routing \cite{ngo2023digital}. As illustrated in Figure~\ref{fig:topology}, which depicts both the physical system and its digital replica, graph-based patterns can be analyzed using underlying GNN models operating within the digital replica \cite{ngo2023digital}. These intelligent models capture interactions between interconnected nodes as they relay traffic through backbone nodes, hub towers and distributed nodes, enabling a comprehensive understanding of hierarchical and spatial dependencies across the interconnected network. Furthermore, the integration of GNNs into digital twin frameworks has shown to enable controlled simulation environments, where topology-driven performance metrics can be learned and evaluated systematically \cite{ngo2023digital}.
\\ While the current state-of-the-art GNN methods have shown remarkable progress, there is still room to exploit HSTGNN especially in mobile networks. Recent works discussed by authors in \cite{Wang2025b}, shows that hybrid spatial-temporal architectures that combine multiple spatial learning strategies with attention-based fusion can effectively capture static, dynamic, and semantic dependencies within complex graph structures, leading to improved predictive accuracy and generalization. Additionally, incorporating hybrid spatial learning with dynamic graph construction and temporal decoupling mechanisms further improves the ability to model evolving network structures and heterogeneous temporal patterns, particularly in long-term forecasting scenarios as demonstrated in the works presented in \cite{Wang2025}.
 Extending this superior capability of HSTGNN into the NDT framework, our work aims to fill a gap in the literature that remains largely unexplored by integrating them to demonstrate their improved performance against the current state-of-the-art GNN approaches. In the subsequent section, we present the methodology of our framework.
\section{Methodology}
\subsection{Dataset and Topology}
With inspiration from \cite{Topologyzoo}, we built our framework on Internet Topology Zoo, a collection of ISP network topologies transcribed into GraphML and GML \cite{afourmy3dzoo} for use in network topology research. The dataset provides a diverse collection of real-world network topologies with varying scales and structural characteristics, enabling robust evaluation across different network conditions. It includes both medium and large-scale topologies, with node counts ranging from tens to several hundreds. This diversity supports the emulation of key 6G properties, particularly heterogeneous network integration and network-of-networks architectural concept \cite{11230693}. As summarized in Table~\ref{tab:topology_dataset}, a combined graph of 1210 nodes and 1572 initial edges was used in our framework, augmented to 4354 edges after preprocessing for GNN input. All features where normalized to $[0, 1]$ with min-max scaling for improved model performance \cite{shantal2023feature}. The dataset split followed a  train/validation/test split as additionally shown in Table~\ref{tab:topology_dataset}, preventing label leakages.
\subsection{Feature Engineering}

The 10-dimensional feature vector $\mathbf{x}_i$ for node $i$ included: degree, betweenness centrality, clustering coefficient, closeness centrality, PageRank, latitude, longitude, 2-hop neighborhood degree, eigenvector centrality, and core number. These features are of optimal importance and adopted in graph-based learning to capture both structural importance and spatial context \cite{tam2022gnn, ngo2023digital}. Table~\ref{tab:features} summarizes the statistical properties of the raw node features prior to min-max normalization.

\begin{table}[t]
\caption{Dataset split and key training configurations}
\label{tab:topology_dataset}
\begin{center}
\setlength{\tabcolsep}{4pt}
\resizebox{\columnwidth}{!}{
\begin{tabular}{ll|ll}
\toprule
\textbf{Parameter} & \textbf{Setting} & \textbf{Parameter} & \textbf{Setting} \\
\midrule
Topology & Internet Topology. Zoo & Optimizer & AdamW \\
Nodes & 1{,}210 & Hidden dim & 96 \\
links & 1{,}572 & Output dim & 2 \\
Train/Val/Test & 847/181/182 & Dropout & 0.20 \\
Input feat. dim & 10 & Random seed & 42 \\
Pred. targets & RTT, Pkt. Loss & & \\
\bottomrule
\end{tabular}}
\end{center}
\end{table}

\begin{table}[t]
\caption{Feature statistics for the 10-dimensional node feature vector}
\label{tab:features}
\begin{center}
\small
\resizebox{\columnwidth}{!}{
\begin{tabular}{lrrrr}
\toprule
\textbf{Feature} & \textbf{Min} & \textbf{Max} & \textbf{Mean} & \textbf{Std} \\
\midrule
Degree & 1.0000 & 30.0000 & 2.5983 & 2.3427 \\
Betweenness Centrality & 0.0000 & 0.4564 & 0.0127 & 0.0402 \\
Clustering Coefficient & 0.0000 & 1.0000 & 0.1074 & 0.2609 \\
Closeness Centrality & 0.0313 & 0.1012 & 0.0638 & 0.0122 \\
PageRank & 0.0001 & 0.0121 & 0.0008 & 0.0010 \\
Latitude & -90.0000 & 90.0000 & 35.9299 & 24.7228 \\
Longitude & -180.0000 & 180.0000 & -10.8449 & 76.2792 \\
2-Hop Deg. Sum & 2.0000 & 82.0000 & 12.2397 & 10.6532 \\
Eigenvector Centrality & 0.0000 & 0.3372 & 0.0025 & 0.0286 \\
Core Number & 1.0000 & 8.0000 & 1.7000 & 0.8192 \\
\bottomrule
\end{tabular}}
\end{center}
\end{table}

\subsection{Target Generation}

The target features (RTT \& packet loss) adopted in our study were generated through a topology-driven simulation procedure. In the implementation, edge attributes such as geographic distance, propagation delay, link capacity, and utilization were derived from the connected network graph. Traffic demand was then emulated between sampled source-destination pairs using weighted shortest paths, allowing the underlying model to reflect topology-aware routing behavior in a controlled and reproducible setting.
For each node, RTT was modeled as a function of path-level propagation delay, queueing delay driven by utilization, and additive noise, which is consistent with work showing that queueing delay dominates total network delay and that latency-based feedback can resolve very fine-grained queuing effects \cite{Moon2015LatencybasedCD}. Packet loss was generated from a congestion-sensitive probability model, which fits the literature showing that packet loss can be bursty at sub-RTT timescales and tightly coupled to queueing behavior \cite{wei2007packet}. This design framework preserves reproducibility while producing targets that are closer to real network behavior.

\section{HSTGNN Architecture}

\subsection{Design Principle}

Rather than choosing a single GNN paradigm, HSTGNN adopts a hybrid spatio-temporal design that fuses static and dynamic graphs and uses a GRU with graph convolution to model temporal evolution and spatial dependencies \cite{zhao2025review}, while combining three complementary message-passing mechanisms in parallel. This hybrid structure matches recent spatial-temporal graph models that aggregate neighbors within each slice of a graph, exchange information across graph slices over time, and combine graph Fourier or spectral convolution operators with attention-based or temporal gated modules. These representations are then refined in stacked layers or blocks, which is consistent with architectures that repeat hierarchical aggregation or block-wise temporal modeling across depth. Additionally, this hybrid design framework follows recent trends demonstrating that combining spatial, spectral, and attention-based GNNs improves representational power and generalization \cite{liang2022survey}.

\subsection{Core Components}

$\mathbf{H}^{(\ell)}$ denoting the hidden representation at layer $\ell$. Each multi-scale block contains three parallel branches:

\textbf{Branch 1 -- Local (GraphSAGE):}
\begin{equation}
\mathbf{h}_{i}^{(\text{sage})} =
\sigma\left(
\mathbf{W}_1
\left[
\mathbf{h}_{i}^{(\ell)} \,\|\, \mathrm{AGG}_{j \in \mathcal{N}(i)} \mathbf{h}_{j}^{(\ell)}
\right]
\right),
\end{equation}
which captures local neighborhood structures through inductive aggregation.

\textbf{Branch 2 -- Spectral (Chebyshev):}
\begin{equation}
\mathbf{h}_{i}^{(\text{cheb})} =
\sum_{k=0}^{K} \theta_k T_k(\tilde{\mathbf{L}})\mathbf{h}_{i}^{(\ell)},
\end{equation}

\textbf{Branch 3 -- Attention (TransformerConv):}
\begin{equation}
\mathbf{h}_{i}^{(\text{attn})} =
\sum_{j \in \mathcal{N}(i)\cup\{i\}}
\alpha_{ij}\mathbf{W}_3\mathbf{h}_{j}^{(\ell)},
\end{equation}

where $\tilde{\mathbf{L}} = \mathbf{I}_N - \mathbf{D}^{-1/2}\mathbf{A}\mathbf{D}^{-1/2}$ is the normalized graph Laplacian and $K=3$ in the proposed model and $\alpha_{ij}$ are learned attention coefficients that adaptively quantify the relative importance of each neighbor.

The branch outputs are concatenated and passed through a learned channel-gating mechanism:
\begin{equation}
\mathbf{h}_{i}^{(\text{cat})} =
\left[
\mathbf{h}_{i}^{(\text{sage})} \,\|\, 
\mathbf{h}_{i}^{(\text{cheb})} \,\|\, 
\mathbf{h}_{i}^{(\text{attn})}
\right],
\end{equation}
\begin{equation}
\mathbf{h}_{i}^{(\ell+1)} =
\phi\left(
\mathbf{g}^{(\ell)} \odot \mathbf{h}_{i}^{(\text{cat})}
+ \mathbf{R}\mathbf{h}_{i}^{(\ell)}
\right),
\end{equation}
where $\mathbf{g}^{(\ell)}$ is a learned channel gate, $\mathbf{R}$ is a residual projection, and $\phi(\cdot)$ denotes normalization and nonlinearity.

The full HSTGNN architecture stacked three such multi-scale blocks. After these blocks, two learnable temporal modulation modules apply channel-wise scale-shift transformations and feed-forward refinement. A final multi-head graph attention layer further refines the learned representation. The output stage combines graph-derived and raw-feature-derived representations through a learned mixing mechanism and adds a direct skip connection from the input features to the prediction head. This design improves expressiveness while maintaining stable optimization.

\subsection{Loss Function and Training}

Training followed a full-graph node-regression setting using the AdamW optimizer, learning-rate scheduling, gradient clipping, and early stopping. The training objective combined Huber loss, mean squared error, and a ramped correlation-based term:
\begin{equation}
\mathcal{L}_{\text{HSTGNN}} =
0.50\,\mathcal{L}_{\text{Huber}}
+ 0.25\,\mathcal{L}_{\text{MSE}}
+ 0.25\,r(t)\,\mathcal{L}_{\text{corr}},
\end{equation}
where $r(t)$ is a ramp factor that gradually increases the influence of the correlation term during training.

For the baseline models, the training objective is:
\begin{equation}
\mathcal{L}_{\text{base}} =
\mathcal{L}_{\text{MSE}} + 0.05\,\mathcal{L}_{\text{L1}}.
\end{equation}

For efficient evaluation, the models were trained and their final performance evaluated on a held-out test set using $R^2$, MAE, RMSE, and Huber loss as key evaluation metrics. The coefficient of determination $R^2$ defined as:
\begin{equation}
R^2 = 1 - \frac{\sum_{i=1}^{n}(y_i - \hat{y}_i)^2}{\sum_{i=1}^{n}(y_i - \bar{y})^2}
\end{equation}
capturing the proportion of variance explained by the model and providing a measure of overall goodness-of-fit \cite{chicco2021coefficient}. Mean Absolute Error computed as:
\begin{equation}
\text{MAE} = \frac{1}{n}\sum_{i=1}^{n}|y_i - \hat{y}_i|
\end{equation}
reflecting the average magnitude of prediction errors and offering an interpretable measure of typical deviation \cite{Chai}. Root Mean Squared Error given by:
\begin{equation}
\text{RMSE} = \sqrt{\frac{1}{n}\sum_{i=1}^{n}(y_i - \hat{y}_i)^2}
\end{equation}
penalizing larger errors more strongly due to its quadratic nature \cite{Chai}. Finally, Huber loss which combines squared-error and absolute-error behavior \cite{meyer2021alternative}:
\begin{equation}
L_\delta(y, \hat{y}) = \frac{1}{n}\sum_{i=1}^{n} \begin{cases} 
\frac{1}{2}(y_i - \hat{y}_i)^2 & \text{if } |y_i - \hat{y}_i| \leq \delta \\ 
\delta\left(|y_i - \hat{y}_i| - \frac{\delta}{2}\right) & \text{otherwise} 
\end{cases}
\end{equation}
making it less sensitive to outliers (with $\delta=0.1$ in our experimentation framework).

\section{Results Discussion and Analysis}

\subsection{Overall Performance}

\begin{figure}[htbp]
\centerline{\includegraphics[width=0.9\columnwidth]{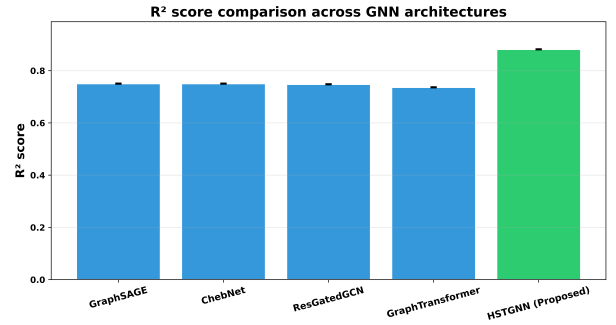}}
\caption{HSTGNN attains the highest R² score ($\approx 0.8816$), outperforming all baseline frameworks.}
\label{fig:r2}
\end{figure}

\begin{figure}[htbp]
\centerline{\includegraphics[width=0.9\columnwidth]{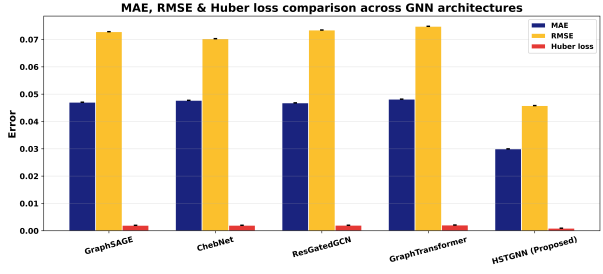}}
\caption{HSTGNN achieves the lowest MAE, RMSE and Huber loss among all compared baselines.}
\label{fig:mae}
\end{figure}

In this subsection and the following ones, we provide an analysis of the results obtained. As demonstrated in Figures~\ref{fig:r2} and ~\ref{fig:mae}, HSTGNN achieved the highest coefficient of determination with $R^2 \approx 0.8816$, while also obtaining the lowest MAE $\approx 0.0300$, the lowest RMSE $\approx 0.0458$, and the lowest Huber loss $\approx 0.00093$. These results show that the proposed hybrid model produces the most accurate and most stable predictions overall.
Among the baseline methods, ChebNet is the strongest with $R^2 \approx 0.7504$, MAE $\approx 0.0477$, and RMSE $\approx 0.0703$, followed closely by GraphSAGE with $R^2 \approx 0.7503$, ResGatedGCN with $R^2 \approx 0.7478$, and GraphTransformer with $R^2 \approx 0.7363$.

The fact that HSTGNN outperforms the rest in all error performance metrics indicates that the framework is not only accurate on average, but is also less prone to large prediction errors. This makes the framework more aligned for practical NDT deployment, where occasional large estimation errors can be operationally costly especially in replicating the Physical Twin (PT).

\begin{figure}[htbp]
\centerline{\includegraphics[width=\columnwidth, height=5cm]{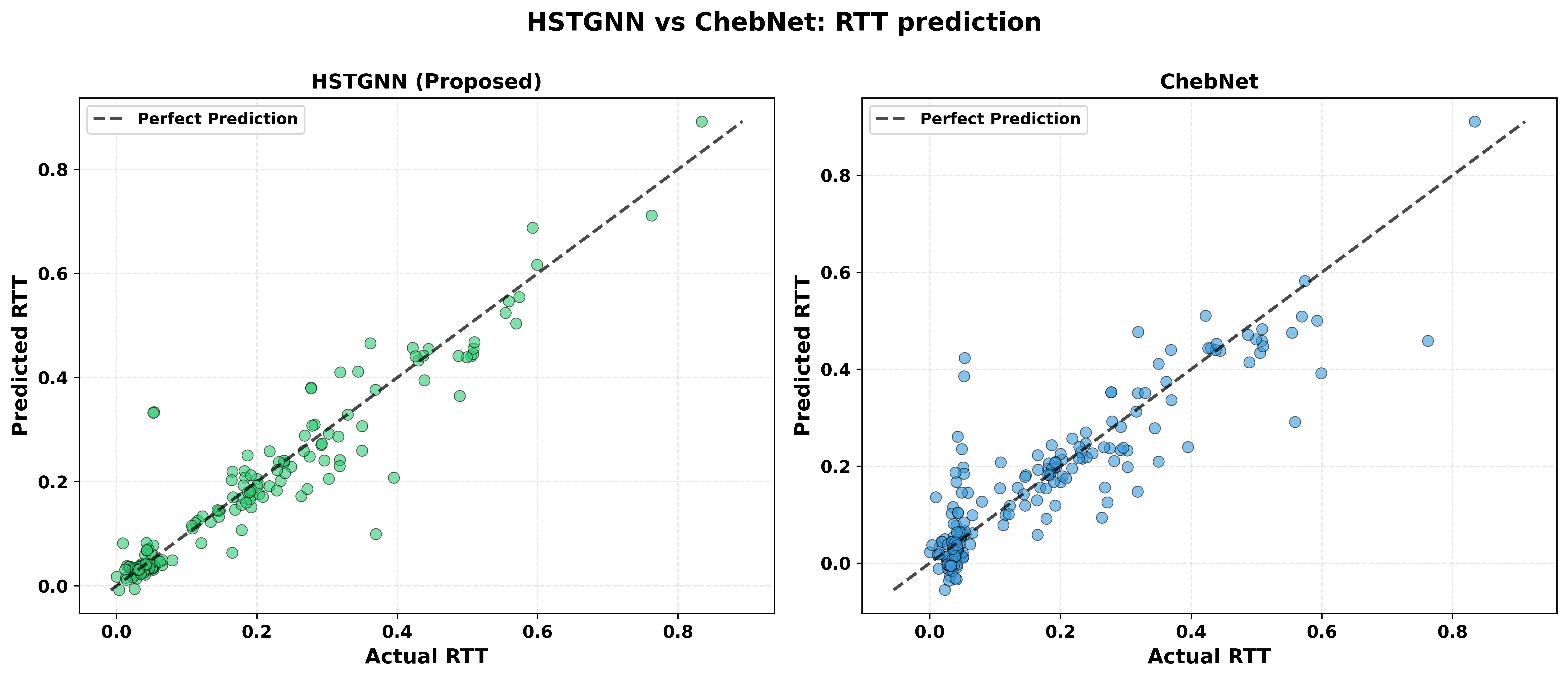}}
\caption{Scatter plot benchmarking HSTGNN against the strongest baseline.}
\label{fig:scater}
\end{figure}

Additionally, the prediction-versus-actual scatter plots in Figure.~\ref{fig:scater} provide additional qualitative evidence of HSTGNN's superiority. For a well-calibrated model, predictions should cluster closely around the diagonal identity line. In the scatter plots, HSTGNN exhibits the tightest concentration around this line compared to the highest performing baseline ChebNet.
This visual pattern is significant because it confirms that the performance gain is not merely numerical but also visual, showing that HSTGNN produces predictions that track the true RTT values more closely across the full output range.

Moreover, Table~\ref{tab:hstgnn_gain} highlights the practical gain of HSTGNN over the strongest baseline. In absolute terms, the model improves $R^2$ by $17.5\%$. In relative terms, it reduces MAE by $37.1\%$ and RMSE by $34.9\%$, while also requiring substantially fewer epochs $\approx 146$ to converge. This is an important result because it shows that the performance advantage of HSTGNN is not achieved at the expense of optimization stability. Instead, the framework is both more accurate and more resource efficient during training.

\begin{table}[t]
\caption{HSTGNN improvement over the strongest baseline (percentage improvement computed as $(B - H)/B \times 100$}
\label{tab:hstgnn_gain}
\centering
\small
\resizebox{\columnwidth}{!}{
\begin{tabular}{lccc}
\toprule
\textbf{Metric} & \textbf{HSTGNN (H)} & \textbf{ChebNet (Best baseline (B))} & \textbf{Absolute improvement} \\
\midrule
$R^2$ & $\mathbf{\uline{0.8816}}$ & 0.7504 & $\mathbf{\uline{17.5\%} \uparrow}$ \\
MAE & $\mathbf{\uline{0.0300}}$ & 0.0477 & $\mathbf{\uline{37.1\%} \downarrow}$ \\
RMSE & $\mathbf{\uline{0.0458}}$ & 0.0703 & $\mathbf{\uline{34.9\%} \downarrow}$ \\
Epochs & $\mathbf{\uline{146}}$ & 278 & $\mathbf{\uline{47.5\%} \downarrow}$ \\
\bottomrule
\end{tabular}}
\end{table}

\subsection{Variance analysis}

\begin{figure}[htbp]
\centerline{\includegraphics[width=\columnwidth,height=4.7cm]{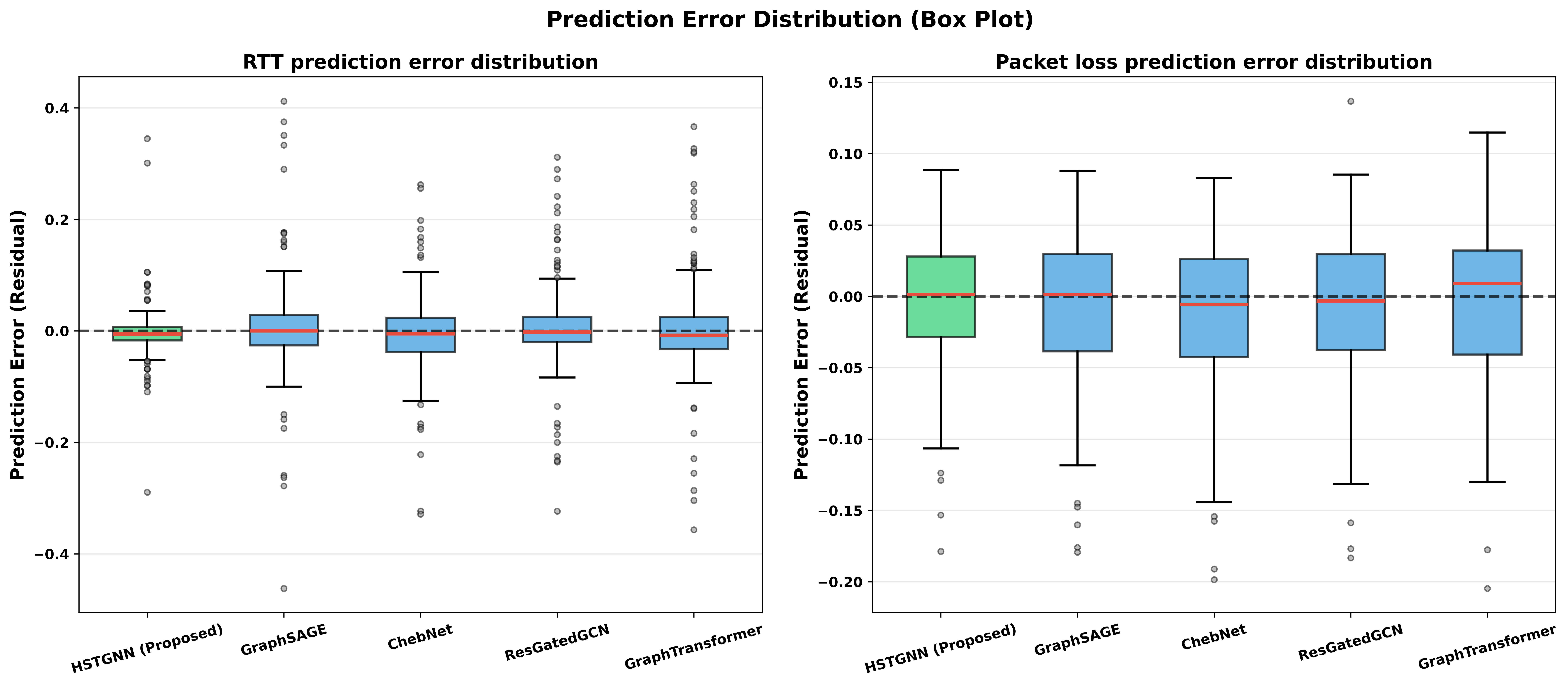}}
\caption{Comparative box plots.}
\label{fig:variace}
\end{figure}

As the authors in \cite{modesto2025towards} suggest, low variance in prediction errors is a critical requirement for reliable Network Digital Twin deployment, because it reflects the consistency and stability of model outputs under inherently stochastic network conditions. In AI driven frameworks, high variability in errors is closely associated with increased uncertainty and reduced trustworthiness of predictions, even when average accuracy appears competitive \cite{hullermeier2021aleatoric}. Therefore, minimizing error variance is essential for ensuring reproducible and dependable model behavior in real-world environments. The presented plots in Figure~\ref{fig:variace} clearly demonstrates that HSTGNN achieves substantially lower error variance than all baseline model frameworks, as evidenced by the tightly concentrated distributions around zero for both RTT and packet loss predictions. Such compact distributions indicate highly consistent predictions across varying conditions, which is essential for operational decision-making in dynamic networks.

\subsection{Operational Relevance for Network Digital Twins}

A NDT is expected to provide reliable predictions of operational metrics before control actions are applied to the real system in a closed loop framework \cite{apostolakis2023digital}. In the results obtained, the proposed model achieves strong performance on both RTT and packet loss jointly, which is significant because these two metrics reflect complementary aspects of service quality that is, latency and reliability.
The results also suggest that HSTGNN is suitable for scenarios involving planning, validation, and what-if case analysis. The lower MAE and RMSE imply more trustworthy prediction outputs, while the higher $R^2$ indicates that the framework captures the main patterns governing network behavior. Together, these properties strengthen the usefulness of the proposed architecture as a predictive framework inside a mobile network digital twin. Moreover, in Table~\ref{tab:req_alignment_summary} we show experimental findings in a broader systems context where we show alignment of these capabilities with the direction of 6G intelligent networking and with 3GPP standadization road-map related to analytics-driven automation, service assurance, and closed-loop control. The experimental evidence indicates that the proposed architecture is well positioned for such future integration.

\begin{table}[t]
\caption{Alignment with NDT, 6G, and 3GPP requirements}
\label{tab:req_alignment_summary}
\begin{center}
\setlength{\tabcolsep}{4pt}
\renewcommand{\arraystretch}{0.85}
\resizebox{\columnwidth}{!}{

\begin{tabular}{p{2cm}p{5.5cm}c}
\toprule
\textbf{Requirement area} & \textbf{Representative requirement} & \textbf{HSTGNN} \\
\midrule
Network Digital Twin & Topology-aware, data-driven KPI prediction for what-if analysis and proactive management~\cite{apostolakis2023digital,Sun2022} & \checkmark \\
6G & AI-native~\cite{Katz2026}, ultra-responsive and intelligent adaptability~\cite{fitzek2026promise} & \checkmark \\
3GPP Rel.~19 & AI/ML lifecycle management, intent-driven enhancements, and NDT studies~\cite{3gpp2024rel18} & \checkmark \\
\bottomrule
\end{tabular}}
\end{center}
\end{table}

\section{conclusion}

Overall, the results show that HSTGNN is not merely a marginal extension of existing GNN baselines, but a stronger modeling framework for topology-aware network digital twins. Its superiority is supported numerically by the best $R^2$, MAE, RMSE, and Huber loss values. The consistency of the results our manuscript presents concludes that hybrid spatial-temporal graph learning is a promising direction for predictive network digital twin design. Future work will focus on developing a closed-loop NDT system that couples HSTGNN predictions with network control algorithms for autonomous optimization necessary in the forthcoming 6G ecosystem.

\section*{Acknowledgment}
This work was partially funded by the MARE project. MARE has received funding from the Smart Networks and Services Joint Undertaking (SNS JU) under the European Union’s Horizon Europe research and innovation programme under Grant Agreement No 101191436.

\bibliographystyle{unsrt}
\bibliography{references}
\end{document}